\documentclass[conference, letterpaper]{IEEEtran}
\pdfoutput=1  %

\usepackage{url}
\usepackage{threeparttable}
\usepackage{multirow}

\usepackage{siunitx}
\DeclareSIUnit{\cycles}{cycles}

\usepackage[%
  activate={true,nocompatibility},%
  final,%
  kerning=true,%
  spacing=true,%
  stretch=20,%
  shrink=25,%
  factor=900,%
  ]{microtype}

\usepackage{algorithm2e}

\SetCommentSty{mycommfont}

\usepackage{graphicx}

\makeatletter
\providecommand*{\input@path}{}
\g@addto@macro\input@path{{src/}}%
\makeatother

\usepackage[acronyms,nohypertypes={acronym},nomain]{glossaries}
\newacronymstyle{custom-acr}{\glsgenacfmt}{

}
\setacronymstyle{custom-acr}

\usepackage{cleveref}
\crefname{algocf}{Algorithm}{Algorithms}
\crefname{section}{\S\!}{\S\!}
\crefname{figure}{Fig.}{Figs.}

\newcommand{\glsknown}[1]{\glsunset{#1}\gls{#1}}
\newcommand{\glsknownpl}[1]{\glsunset{#1}\glspl{#1}}
\newcommand{\glsunknown}[1]{\glsreset{#1}\gls{#1}}
\newcommand{\Glsunknown}[1]{\glsreset{#1}\Gls{#1}}
\newcommand{\glsunknownpl}[1]{\glsreset{#1}\glspl{#1}}

\newcommand{\axi}{AXI\nobreak\hspace{.125em}4}

\usepackage{booktabs} %

\newcommand\blfootnote[1]{%
  \begingroup
    \renewcommand\thefootnote{}\footnote{#1}%
    \addtocounter{footnote}{-1}%
  \endgroup
}

\usepackage{tikz}
\newcommand{\copyrighttext}{%
  \footnotesize%
  \textcopyright{}~2018 IEEE.
  Personal use of this material is permitted.
  Permission from IEEE must be obtained for all other uses.%

  Accepted for publication in \textit{Proceedings of the 36th IEEE International Conference on Computer Design (ICCD)}, October 7-10, 2018, Orlando, FL, USA.
}
\newcommand{\copyrightnotice}{%
  \vspace*{-2\baselineskip}
  \begin{tikzpicture}[remember picture, overlay]%
    \node[anchor=south, yshift=15pt] at (current page.south) {%
      \fbox{%
        \parbox{\dimexpr\textwidth-\fboxsep-\fboxrule\relax}{\copyrighttext}%
      }%
    };%
  \end{tikzpicture}%
}

\begin{document}

\bstctlcite{IEEEexample:BSTcontrol}\bstctlcite{IEEEexample:BSTcontrol}

\title{Scalable and Efficient Virtual Memory Sharing in Heterogeneous SoCs with TLB Prefetching and MMU-Aware DMA Engine}

\author{%
  \IEEEauthorblockN{%
    Andreas Kurth\IEEEauthorrefmark{1},
    Pirmin Vogel\IEEEauthorrefmark{1},
    Andrea Marongiu\IEEEauthorrefmark{2}\IEEEauthorrefmark{1}, and
    Luca Benini\IEEEauthorrefmark{1}\IEEEauthorrefmark{2}%
  }%
  \IEEEauthorblockA{%
    \IEEEauthorrefmark{1}%
      Integrated Systems Laboratory, ETH Zurich, Switzerland\\%
      \texttt{lastname@iis.ee.ethz.ch}%
    \\%
    \IEEEauthorrefmark{2}%
      University of Bologna, Italy\\%
  }%
}

\newacronym{acp}{ACP}{Accelerator Coherency Port}
\newacronym{api}{API}{application programming interface}
\newacronym{ast}{AST}{abstract syntax tree}
\newacronym{axi}{AXI}{Advanced eXtensible Interface}\glsunset{axi}

\newacronym{cma}{CMA}{Contiguous Memory Allocator}
\newacronym{cpu}{CPU}{central processing unit}\glsunset{cpu}
\newacronym{cu}{CU}{compute unit}

\newacronym{ddg}{DDG}{data dependency graph}
\newacronym{dma}{DMA}{direct memory access}
\newacronym{dram}{DRAM}{dynamic random access memory}
\newacronym{dvm}{DVM}{distributed virtual memory}

\newacronym{fd}{FD}{face detection}
\newacronym{fifo}{FIFO}{first-in-first-out}\glsunset{fifo}
\newacronym{fpga}{FPGA}{field-programmable gate array}\glsunset{fpga}
\newacronym{fpu}{FPU}{floating-point unit}

\newacronym{gpu}{GPU}{graphics processing unit}\glsunset{gpu}
\newacronym{gpgpu}{GPGPU}{general purpose GPU}

\newacronym{hal}{HAL}{hardware abstraction layer}
\newacronym[longplural={heterogeneous systems on chip}]{hesoc}{HeSoC}{heterogeneous system on chip}
\newacronym{hw}{HW}{hardware}\glsunset{hw}

\newacronym{iommu}{IOMMU}{input/output memory management unit}
\newacronym{iotlb}{IOTLB}{input/output translation lookaside buffer}
\newacronym{isa}{ISA}{instruction set architecture}

\newacronym[shortplural={LDSes}]{lds}{LDS}{linked data structure}
\newacronym{lpae}{LPAE}{Large Physical Address Extension}
\newacronym{lru}{LRU}{least recently used}

\newacronym{mac}{MAC}{multiply-accumulate operation}
\newacronym{mc}{MC}{Memory Copy}
\newacronym{mh}{MH}{miss handling}
\newacronym{mht}{MHT}{miss handling thread}
\newacronym{mimd}{MIMD}{multiple instruction multiple data}
\newacronym{mmu}{MMU}{memory management unit}

\newacronym[shortplural={OSes}]{os}{OS}{operating system}

\newacronym{pc}{PC}{Pointer Chasing}
\newacronym{pe}{PE}{processing element}
\newacronym{pgd}{PGD}{Page Global Directory}
\newacronym{pht}{PHT}{Prefetching Helper Thread}
\newacronym{pmca}{PMCA}{programmable many-core accelerator}
\newacronym{pmd}{PMD}{Page Middle Directory}
\newacronym{pr}{PR}{Page\-Rank}
\newacronym{pte}{PTE}{Page Table Entry}
\newacronym{ptw}{PTW}{page table walker}
\newacronym{pt}{PT}{page table}
\newacronym{pud}{PUD}{Page Upper Directory}

\newacronym{rab}{RAB}{Remapping Address Block}
\newacronym{rft}{RFT}{Random Forest Traversal}
\newacronym{rhf}{RHF}{Random Hough Forest}

\newacronym[longplural={systems on chip}]{soc}{SoC}{system on chip}
\newacronym[longplural={scratchpad memories}]{spm}{SPM}{scratchpad memory}
\newacronym{spmd}{SPMD}{single program multiple data}
\newacronym{svm}{SVM}{shared virtual memory}
\newacronym{soa}{SoA}{state of the art}
\newacronym{sw}{SW}{software}\glsunset{sw}

\newacronym{tlb}{TLB}{translation lookaside buffer}

\newacronym{va}{VA}{virtual address}
\newacronym{vmm}{VMM}{virtual memory management}

\newacronym{wt}{WT}{Worker Thread}

\maketitle
\copyrightnotice

\begin{abstract}
\Glsunknown{svm} is key in heterogeneous \glsunknownpl{soc}, which combine a general-purpose host processor with a many-core accelerator, both for programmability and to avoid data duplication.
However, \gls{svm} can bring a significant run time overhead when \glsunknown{tlb} entries are missing.
Moreover, allowing \glsknown{dma} burst transfers to write \gls{svm} traditionally requires buffers to absorb transfers that miss in the \gls{tlb}.
These buffers have to be overprovisioned for the maximum burst size, wasting precious on-chip memory, and stall all \gls{svm} accesses once they are full, hampering the scalability of parallel accelerators.

In this work, we present our \gls{svm} solution that avoids the majority of \gls{tlb} misses with prefetching, supports parallel burst \gls{dma} transfers without additional buffers, and can be scaled with the workload and number of parallel processors.
Our solution is based on three novel concepts:
To minimize the rate of \gls{tlb} misses, the \gls{tlb} is proactively filled by compiler-generated Prefetching Helper Threads, which use run-time information to issue timely prefetches.
To reduce the latency of \gls{tlb} misses, misses are handled by a variable number of parallel Miss Handling Helper Threads.
To support parallel burst \gls{dma} transfers to \gls{svm} without additional buffers, we add lightweight hardware to a standard \gls{dma} engine to detect and react to \gls{tlb} misses.
Compared to the state of the art, our work improves accelerator performance for memory-intensive kernels by up to 4$\times$ and by up to \SI{60}{\percent} for irregular and regular memory access patterns, respectively.\blfootnote{This work was partially funded by the EU's H2020 projects HERCULES (No.~688860) and OPRECOMP (No.~732631).}
\end{abstract}

\section{Introduction}
\label{sec:introduction}

\glsresetall{}

In \glspl{hesoc}, a general-purpose multicore \glsknown{cpu}, the \emph{host}, is co-integrated with \glspl{pmca} on a single die.
This design holds the promise of combining the versatility of the host \gls{cpu} with the energy efficiency and computing performance of the highly parallel accelerators.

One of the major difficulties in programming \glspl{hesoc} is having to explicitly manage the multi-level, non-uniform memory system.
On the host, coherent caches and virtual addresses make the memory hierarchy completely transparent to the application programmer.
On the \gls{pmca}, however, \glspl{spm} are often preferred to hardware-managed caches for the implementation of on-chip memory hierarchies.
\Glspl{spm} are physically addressed and data transfers to and from them are controlled by software, preferably using \gls{dma} transfers.

To alleviate this difficulty and to enable sharing of linked data structures, the Heterogeneous System Architecture Foundation~\cite{HSA} pushed an architectural model where host and \glspl{pmca} communicate via coherent \gls{svm}.
For coherency with the host data caches, most \glsknownpl{soc} today offer accelerators access to coherent interconnects~\cite{stuechli_capi,goodacre_acp}.
For the translation of virtual addresses, there are two main approaches:
In the all-hardware approach followed by many embedded \gls{soc} vendors, a full-fledged \gls{iommu}\glsunset{mmu} translates addresses autonomously~\cite{pichai_gpu_address_translation_2014,arm_mmu_500}.
It is comprised of a \gls{tlb}, parallel hardware \glspl{ptw}, transaction and data buffers, and coherent page table caches.
The alternative is a hybrid hardware-software design, which consists of a \gls{tlb} controlled by software (e.g., by a kernel driver on the host~\cite{lavasani_fpga_2014,vogel_lightweight_2016} or directly by the accelerator~\cite{vogel_on_acc_ptw_2017}).
We subsequently refer to the former class of \glspl{iommu} as \emph{conventional} and to the second class as \emph{hybrid}.
While conventional \glspl{iommu} have the advantage of being transparent to the \glspl{pmca} and of offering the minimal latency for handling an isolated \gls{tlb} miss, they have three significant drawbacks:

First, parallel, interleaved accesses by \glspl{pmca} to independent virtual addresses require parallel \glspl{ptw}.
While the number of parallel accesses is a time-variant run-time property of programs executed by the \glspl{pmca}, the number of \glspl{ptw} is a fixed design parameter in conventional \glspl{iommu}.
To accommodate a wide range of parallel workloads, the number of parallel hardware \glspl{ptw} must be overprovisioned, wasting hardware resources in most use cases.

Second, enabling \gls{dma} engines to access \gls{svm} in conventional \glspl{iommu} requires a data buffer in the \gls{iommu} to absorb write bursts to addresses that miss in the \gls{tlb}.
This buffer requires a significant amount of memory: it must have at least the size of the largest \gls{dma} burst, and is usually even larger because no more \gls{svm} accesses (by \emph{any} master, not just the missing \gls{dma} engine) can be processed once (and as long as) that buffer is full. %

Third, a conventional \gls{iommu} manages its \gls{tlb} purely reactively: new entries are set up only after a \gls{tlb} miss.
While some \glspl{iommu}~\cite{vesely_svm_hesocs,arm_mmu_500} can speculate on future memory accesses based on past access patterns, doing so is very inaccurate for nonlinear, interleaved access patterns and negatively affects performance~\cite{vesely_svm_hesocs}.
Misses can thus occur frequently, and high-performance conventional \glspl{iommu} include coherent data caches for page table entries~\cite{arm_mmu_500} to reduce the latency of handling a \gls{tlb} miss.
If the \gls{tlb} was managed by software threads in the \gls{pmca} instead (as in hybrid \glspl{iommu}), it would be possible for those threads to set up \gls{tlb} entries ahead of time based on run-time information inside the \gls{pmca}.
Research on data caches~\cite{roth_jpp_1999,alsukhni_cdcap_2003} has long shown that prefetchers that know the running program and its status can be far more accurate than those that can only see a stream of memory addresses.

The hybrid design, on the other hand, theoretically does not have these drawbacks.
However, the state-of-the-art implementation~\cite{vogel_on_acc_ptw_2017} features only a single \gls{ptw} thread, only supports \gls{dma} bursts that are guaranteed not to miss (e.g., by locking the corresponding \gls{tlb} entries), and does not perform any prefetching.

In this work, we resolve these limitations.
To the best of our knowledge, this work is the first to:
\begin{enumerate}
  \item implement accurate, compiler-generated prefetching for a shared \gls{tlb} (\cref{sec:impl:prefetching}), which significantly reduces the rate of \gls{tlb} misses,
  \item offer a flexible number of parallel \gls{tlb} miss handlers (\cref{sec:impl:par_mht}), which keeps the miss handling latency constant for scalable parallel workloads, and
  \item offer shared virtual memory accessible by \gls{dma} transfers without additional buffers (\cref{sec:impl:vmm-capable_dma}).
\end{enumerate}

Compared to the state-of-the-art hybrid \gls{iommu}~\cite{vogel_on_acc_ptw_2017}, our contributions improve the \gls{pmca} performance for memory-intensive kernels by up to 4$\times$ and by up to \SI{60}{\percent} for irregular and regular memory access patterns, respectively (\cref{sec:res:synthetic_benchmarks}).
Compared to using data buffers to absorb bursts from \gls{dma} engines, our solution requires two orders of magnitude less memory (\cref{sec:res:dma_hw_req}) and scales better, as it only stalls the missing \gls{dma} engine.

\section{Related Work}
\label{sec:related_work}

The vast majority of commercial systems today features conventional \glspl{iommu}~\cite{arm_mmu_500,intel_gen9,kornaros_io_2014,xilinx_zynqmpsoc_overview} to completely abstract the \gls{svm} implementation from the \glspl{pmca}.
While simple to use, that approach is limited in scalability (handling parallel misses and absorbing burst transfers) and efficiency (reactive \acrshort{tlb} management).

\paragraph{Parallel \acrshort{tlb} miss handling and page table walking}
The fixed number of shared hardware \glspl{ptw} puts an upper bound on the scalability of conventional \glspl{iommu} to parallel accelerators.
A recent study~\cite{vesely_svm_hesocs} has shown that an integrated \glsknown{gpu} with 8 parallel \glspl{cu} quickly saturates the miss handling capabilities of an \gls{iommu} with 16 hardware \glspl{ptw}, after which the \gls{gpu}'s performance becomes bounded by \gls{tlb} miss handling latency.
To avoid this, the current proposal for address translation on \glspl{gpu}~\cite{power_gpu_address_translation_2014,pichai_gpu_address_translation_2014} is to add one \glsknown{mmu} before the L1 cache in every \gls{cu}.
Each such \gls{cu} \gls{mmu} has its private \gls{tlb} and either has its own \gls{ptw}~\cite{pichai_gpu_address_translation_2014} or shares a highly-threaded \gls{ptw}~\cite{power_gpu_address_translation_2014}.
As this approach adds a significant amount of hardware, its parameters (e.g., \gls{tlb} size, number of \glspl{ptw}) must be carefully balanced \emph{at hardware design time} to neither present a bottleneck for \gls{svm}-heavy applications nor reduce the compute-per-area ratio for applications that use \gls{svm} in a lighter way.
The miss handling throughput of hybrid \glspl{iommu}, where page table walks are performed by software threads, on the other hand, can be scaled \emph{at run time} by scheduling \glspl{ptw} when required.
However, efficiently managing a \gls{tlb} shared by many parallel \glspl{pe} and notifying individual \gls{pe} with low latency about handled misses is not trivial.
For this reason, current hybrid \gls{svm} solutions~\cite{vogel_on_acc_ptw_2017,lavasani_fpga_2014} only feature a single \gls{ptw} thread.
In this work, we show how to efficiently manage a shared \gls{tlb} with multiple software \gls{ptw} threads.

\paragraph{Handling bursts missing in the \acrshort{tlb}}
The buffers in conventional \glspl{iommu} that absorb write bursts missing in the \gls{tlb}~\cite{arm_mmu_500} are another limiting factor for accelerators based on \gls{dma} transfers:
When (and as long as) the limit on outstanding misses is reached, the \gls{iommu} cannot translate any further transactions, even if they would hit.
This creates backpressure from the \gls{iommu} slave port to the connected master ports, stalling each master port on its next \gls{svm} access.
Hybrid \glspl{iommu}, on the other hand, signal the \gls{tlb} miss back to the master and drop the transaction~\cite{vogel_lightweight_2016}.
This allows to handle misses on a shared \gls{tlb} in a much more scalable way: instead of creating congestion on shared resources (e.g., buffers in the \gls{iommu}, interconnect), the transaction that missed stays in the source memory, keeping shared resources clear for other accesses.
To support this, the \gls{dma} engine must be able to keep track of bursts that missed and reissue them when the miss has been handled.
In this work, we introduce a lightweight hardware extension that adds this feature for a standard \gls{dma} engine.

\paragraph{Reducing \acrshort{tlb} misses through prefetching}
Reducing the number of \gls{tlb} misses is another effective way to reduce the run time overhead of \gls{svm}, orthogonal to reducing the miss handling latency.
There are two independent strategies to achieve this:
The first is to increase the capacity of the \gls{tlb}, for which both conventional and hybrid \glspl{iommu} feature multi-level \glspl{tlb}~\cite{arm_mmu_500,vogel_lightweight_2016}.
The second is to ensure the timelineness of \gls{tlb} entries, e.g.\ through prefetching.
Prefetching for shared \glspl{tlb} is not yet well-understood:
Some conventional \glspl{iommu} feature a very simple prefetcher, which adds two subsequent pages to the \gls{tlb} in case of a miss to the first~\cite{arm_mmu_500}.
However, this prefetcher is deactivated by default because it harms performance in most cases~\cite{arm_mmu_500}.
Prefetching is also supported by the PCIe Address Translation Services~\cite{pcie_ats}, but a recent study~\cite{vesely_svm_hesocs} examined a benchmark with low locality, found that having a \gls{gpu} prefetch eight contiguous pages degrades performance by up to 3$\times$, and concluded that research on application-aware prefetching is required.
In this work, we design and implement accurate prefetching for a shared \gls{tlb}, which significantly reduces the rate of \gls{tlb} misses.
We focus on \glspl{lds}, which are the predominant source of scattered memory accesses in many programs.
Our design is inspired by the following prefetchers for data caches.

Prefetchers that get information about the running program from software~\cite{roth_jpp_1999,luk_prefetching_1999,karlsson_pa_2000,alsukhni_cdcap_2003,ganusov_edht_2006,ebrahimi_ecdp_2009,lee_pht_multiproc_2009,son_compiler_prefetching_cmps_2009} are far more effective than heuristic hardware units~\cite{cooksey_cdp_2002,collins_pcap_2002} for \glspl{lds}:
Heuristic hardware prefetchers for \glspl{lds} identify pointers as they are loaded from memory, prefetch their content before they are dereferenced, and store it in a separate pointer cache~\cite{collins_pcap_2002} or in the data cache of the processor~\cite{cooksey_cdp_2002}.
All pointers identified by the heuristic hardware are prefetched recursively, which leads to a low prefetch accuracy, consequently polluting the cache and decreasing performance in many cases.
To improve prefetch accuracy, the hardware prefetcher in hybrid hardware/software prefetchers~\cite{ebrahimi_ecdp_2009,alsukhni_cdcap_2003} is controlled by software, e.g., from the main processor through special instructions to identify useful prefetches~\cite{ebrahimi_ecdp_2009} or by running a separate prefetching program~\cite{alsukhni_cdcap_2003}.
Prefetch code can be written manually by a developer~\cite{roth_jpp_1999,ganusov_edht_2006} or generated automatically by a compiler~\cite{karlsson_pa_2000,luk_prefetching_1999,lee_pht_multiproc_2009,son_compiler_prefetching_cmps_2009}, through static or dynamic profiling or both.
Compilers can accurately identify pointers and prioritize their prefetching according to dependencies.
Once prefetchers for \glspl{lds} are accurate, their effectiveness is limited by memory latency, as prefetch targets in \glspl{lds} depend on earlier pointer dereferences.
As a dedicated hardware prefetcher co-located with the processor has the same memory latency as the processor itself, pure software prefetchers have been explored instead~\cite{ganusov_edht_2006,luk_prefetching_1999,lee_pht_multiproc_2009,son_compiler_prefetching_cmps_2009}.
Prefetches inserted inline with the actual program code~\cite{luk_prefetching_1999}, however, are limited to targets that are known when that line of code is executed.
Otherwise, the actual program has to be stalled while the pointers leading to the prefetch target are followed.
A promising alternative is to run an additional prefetching thread on the same multithreaded processor core~\cite{luk_smt_preexec_2001,ganusov_edht_2006} or on another core in the same processor~\cite{lee_pht_multiproc_2009,son_compiler_prefetching_cmps_2009}.
Another important advantage of these separate prefetcher threads is that their throughput can be scaled to the demands of the application at compile-time or at run-time or both, especially for the high degree of parallelism offered by \glspl{pmca}.
A key difficulty of software prefetches executed by another thread is the timeliness of the prefetches, which is why prefetching threads have primarily been explored for coarse-grained prefetching~\cite{son_compiler_prefetching_cmps_2009,lee_pht_multiproc_2009}.

While there are a number of works using heuristic hardware units for \gls{tlb} prefetching~\cite{saulsbury_rtp_2000,kandiraju_tlb_prefetching_2002,lustig_tlb_cmps_2013}, this work (to the best of our knowledge) is the first to use compiler-generated software threads for \gls{tlb} prefetching.
For the first time, this allows to accurately prefetch \gls{tlb} entries for \glspl{lds}.
Compared to the related compiler-generated software prefetchers, our solution is novel in how it issues fine-grained, timely prefetches into a shared resource (the \gls{tlb}) in a scalable way without causing negative interference.

\section{Target Architecture Template}
\label{sec:target_architecture_template}

\begin{figure}
  \centering
  \includegraphics[width=\columnwidth]{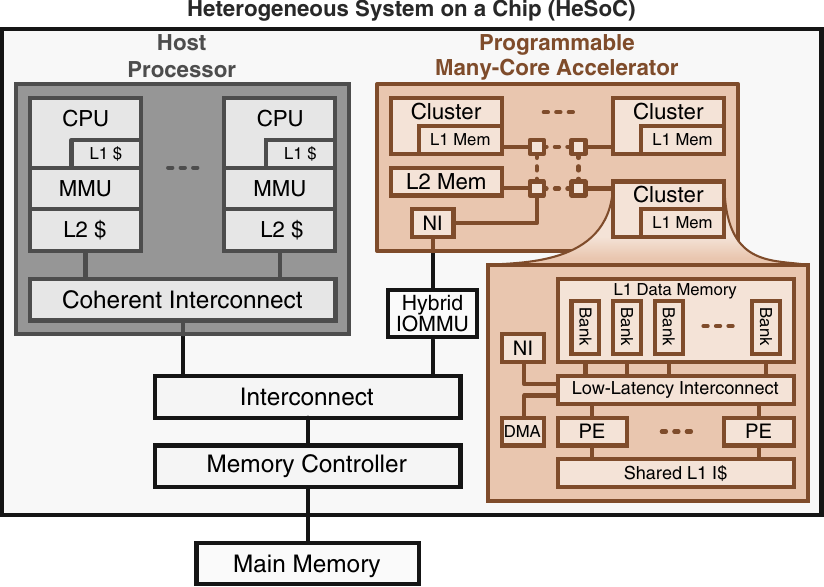}
  \caption{Template of the target architecture.}%
  \label{fig:target_architecture_template}
\end{figure}

The heterogeneous system targeted in this work combines two architecturally different processors in a single chip.
As shown in \cref{fig:target_architecture_template}, the \gls{hesoc} is composed of a general-purpose multi-core \gls{cpu} (the host) and a domain-specific \gls{pmca}.
The host \gls{cpu} features a cache-coherent memory hierarchy, runs a full-fledged operating system, and manages inputs and outputs of the \gls{hesoc}.
The \gls{pmca} complements the host by offering high computational performance and efficiency for specific application domains.

The \gls{pmca} we consider uses a multi-cluster design~\cite{melpignano_platform_2012,kalray} to enable architectural scaling.
In each cluster~\cite{rossi_pulp_cluster_2017}, multiple \glspl{pe} share an L1 data \gls{spm} and an L1 instruction cache~\cite{loi_icache_2018}, both multi-banked (twice as many banks as \glspl{pe}), through a low-latency logarithmic interconnect~\cite{loi_l1icache_2015}.
Multiple clusters are attached to the main network of the \gls{pmca}, through which they share an L2 \gls{spm}.

The off-chip main \gls{dram} is physically shared by the host and the \gls{pmca}.
To exploit data locality, both host and \gls{pmca} keep the most frequently accessed data in fast, local storage of their internal memory hierarchy.
While the host relies on hardware-managed caches, the \gls{pmca} uses multi-channel, high-bandwidth \gls{dma} engines and double-buffering schemes to overlap data movement with computation on data in the L1 \glspl{spm}.
The widely-adopted \axi{} protocol is used between host and \gls{pmca} and on the network in the \gls{pmca}.

The \gls{iommu} allows the \gls{pmca} to share the virtual memory space of an application running on the host.
It is a hybrid design~\cite{vogel_lightweight_2016}, consisting of a \gls{tlb} of configurable size completely managed by software running on the \gls{pmca}.
The \gls{tlb} is fully associative and processes look-ups within a single clock cycle.
In case of a \gls{tlb} miss, the \gls{iommu} stores the metadata in a hardware queue, responds with an error, drops the transaction, and processes the next one.
In case of a \gls{tlb} hit, the \gls{iommu} translates the virtual address to a physical one, forwards the transaction through the master port, and processes the next transaction.

One \gls{pe} of the \gls{pmca} manages the \gls{tlb} in the \gls{iommu}.
Upon a \gls{tlb} miss, it reads the metadata of the missing transaction from the hardware queue in the \gls{iommu}, walks the page table of the offloaded process, replaces an older \gls{tlb} entry with the new translation, and notifies the \gls{pe} that encountered the miss, which then retries the memory access.
As the memory access latency in page table walks dominates the miss handling latency, this software \gls{ptw} has about the same latency as a dedicated hardware \gls{ptw}~\cite{vogel_on_acc_ptw_2017}.

\Glspl{pe} within a cluster execute in a \gls{spmd} fashion and share a multi-ported, multi-banked instruction cache~\cite{loi_l1icache_2015}.
They can exchange data with low latency and low congestion through the L1 data memory, which also offers an atomic \emph{test-and-set} read-modify-write operation.
A dedicated event unit within the cluster supports interrupts, barriers, and software-triggered events for low-overhead synchronization.

Each cluster includes a \gls{dma} engine optimized both in throughput and area for transfers from or to the cluster's tightly-coupled \glspl{spm}~\cite{rossi_dma_2014}.
It supports up to 16 outstanding \axi{} bursts with only minimal internal buffers thanks to the low-latency connection to the \glspl{spm}.
Each \gls{pe} has a private command interface on the \gls{dma} engine, which allows multiple \glspl{pe} to simultaneously enqueue \gls{dma} transfers without the need for synchronization.
The control unit of the \gls{dma} engine internally arbitrates between the per-\gls{pe} command interfaces.
\Glspl{pe} can enqueue coarse-grained transfer commands (up to \SI{64}{\kibi\byte}), which are split up by the control unit into fine-grained bursts (up to \SI{2}{\kibi\byte}) to meet alignment requirements and to facilitate time-multiplexing of downstream \glsknown{axi} resources.
As soon as a coarse-grained transfer is complete, the \gls{dma} engine notifies the \gls{pe} via the event unit.

\section{Implementation}
\label{sec:implementation}

In this section, we detail our compiler-generated \gls{tlb} prefetchers, which significantly reduce the rate of \gls{tlb} misses~(\cref{sec:impl:prefetching}), our scalable multi-threaded \gls{tlb} miss handlers, which keep the \gls{tlb} miss handling latency constant for scalable parallel workloads~(\cref{sec:impl:par_mht}), and our hybrid-\gls{iommu}-capable \gls{dma} engine, which can handle \gls{tlb} misses without additional data buffers~(\cref{sec:impl:vmm-capable_dma}).

\subsection{Helper Thread Prefetching}
\label{sec:impl:prefetching}

As \gls{tlb} miss handlers are dominated by memory latency, frequent \gls{tlb} misses inevitably entail a large run time overhead.
As a consequence, managing the \gls{tlb} solely by reacting on misses is not sufficient.
Instead, \gls{tlb} entries could be set up ahead of the instant they are used in a prefetching manner.

The following observations motivate our prefetcher design:
\begin{itemize}
  \item It shall not rely solely on run-time information (e.g., memory content, memory access patterns).
    This is the black-box approach taken by hardware prefetchers, which is not accurate for \glspl{lds}.
  \item It shall not rely solely on compile-time information (e.g., algorithms, data structures) because this neglects all dynamic information (e.g., data-dependent memory accesses, delays due to interference) required for timely prefetches.
  \item It shall be portable across applications.
    While software prefetches can be inserted manually, doing so effectively requires in-depth knowledge of the target platform and laborious analysis of the application.
    Prefetch insertion shall be fully automatic, not burdening developers.
  \item It shall exploit the cluster architecture of the \gls{pmca}, where tightly-coupled L1 \gls{spm} allows to share the state of \glspl{pe} with low latency and little interference.
  \item Its prefetching throughput shall be scalable at compile-time (because different programs have different memory access patterns and intensities) and at run-time (due to phase-based program behavior~\cite{sherwood_program_phases_2003}).
\end{itemize}

\begin{figure}
  \centering
  \includegraphics[width=\columnwidth]{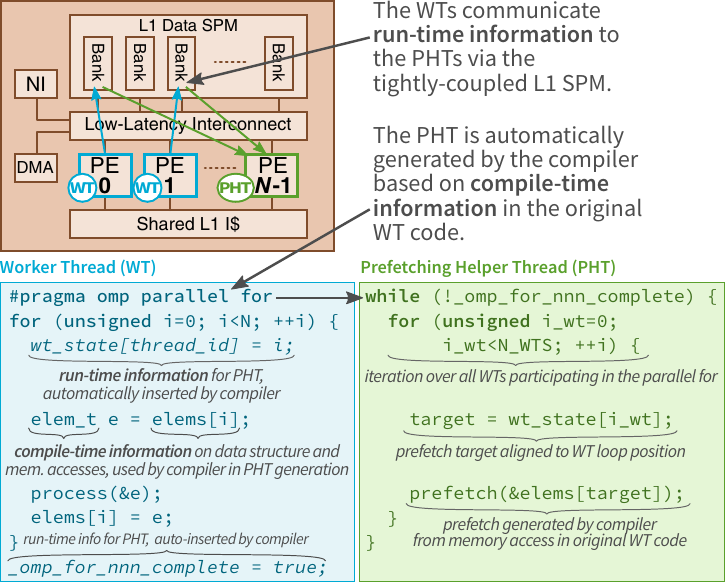}
  \caption{The concept of prefetching with closely-coupled helper threads using the example of a very simple parallel \texttt{for} loop.}%
  \label{fig:helper_thread_prefetching}
\end{figure}

Combined, these considerations led us to the concept of prefetching with closely-coupled helper threads shown in \cref{fig:helper_thread_prefetching}.
Our execution model assumes that part of the \glspl{pe} in a \gls{pmca} cluster are statically allocated to executing the original application workload.
The workload is distributed among as many \emph{\glsunknownpl{wt}} according to the semantics of parallel programming models such as OpenMP~\cite{marongiu_openmp_offload_2015}.
The remaining \glspl{pe} in the cluster are statically allocated to execute \emph{\glsunknownpl{pht}}, which our compiler automatically generates by stripping down the code of the \glspl{wt}:
The idea is to remove all statements that do not access \gls{svm} and do not determine the address or the occurrence of an \gls{svm} access, and to replace \gls{svm} accesses in the remaining code with a call to a prefetch method.
Additionally, the compiler inserts store instructions to the L1 \gls{spm} into the \gls{wt} to share its state of execution and load instructions into the \gls{pht} to read the execution state.
The prefetch method does not modify the \gls{tlb} itself; rather, it checks if a page is currently in the \gls{tlb} and, if it is not, informs the standard \gls{tlb} miss handlers (through the queue of \gls{tlb} misses) that a \gls{tlb} entry must be set up ahead of the moment when a worker thread actually requires the data on a page.
Prefetching \gls{tlb} accesses are described in more detail in~\cref{sec:pf_hw_reqs}.

Prefetches are issued conditionally on the current position of the \gls{wt} relative to the current position of the \gls{pht}.
For example, a fixed window wherein prefetches are issued can be defined:
Assume $w\sb{k}$ is the position of \gls{wt} $k$ in a parallel loop and $d$ and $D$ are the minimum and maximum prefetching distances, respectively.
Then the \gls{pht} has to make sure that the position of its next prefetch for \gls{wt} $k$, $p\sb{k}$, fulfills $w\sb{k} + d \leq p\sb{k} \leq w\sb{k} + D$.
If $p\sb{k} > w\sb{k} + D$, then the \gls{pht} is ahead of \gls{wt} $k$ by more than the maximum prefetching distance and the \gls{pht} will not issue a prefetch.
If $p\sb{k} < w\sb{k} + d$, then the \gls{pht} is behind the minimum prefetching distance and the \gls{pht} will set $p\sb{k}$ to a position inside the window.
When $p\sb{k}$ is inside the window, the \gls{pht} will prefetch at $p\sb{k}$ and then increment $p\sb{k}$.

\subsubsection{Compiler Algorithm to Generate PHTs}%
\label{sec:impl:pht_gen_algo}

The compiler algorithm to generate a \gls{pht} from the code for a \gls{wt} comprises two stages:
The first stage recursively traverses the \gls{ast} of the body compound of the \gls{wt}.
In a forward pass, a \gls{ddg} for each \gls{svm} variable (i.e., a variable that dereferences a pointer to \gls{svm}, possibly through other variables and pointers) is constructed.
In a backward pass, memory accesses to \gls{ddg} leaf nodes are rewritten as prefetches and all statements that are not in the \gls{ddg} of an \gls{svm} variable are removed.
The second stage recursively traverses the modified \gls{ast} and prunes redundant prefetches.

The forward pass is recursively invoked on an \gls{ast} node and scope tuple, where the scope is a list of variables together with their \gls{ddg}.
It creates the \gls{pht} for the given \gls{ast} node by constructing a \gls{ddg} for each \gls{svm} variable and invoking the backward pass afterwards.
The forward pass essentially distinguishes two classes of \gls{ast} nodes:
Declarations extend the scope of the subsequent nodes, and assignments modify the \gls{ddg} of their left-hand side variable.
Compounds establish a local scope, within which the children of the compound are first modified in forward order by the forward pass, then in backward order by the backward pass.

The backward pass is also invoked on an \gls{ast} node and scope tuple.
Based on the \gls{ddg} of each variable, it rewrites dereferences of \gls{svm} pointers that are leaf nodes into prefetches and removes all statements that are not in the \gls{ddg} of an \gls{svm} variable.
It distinguishes three classes of \gls{ast} nodes:
Conditionals and loops add a control flow dependency to variables they reference.
Declarations remove the declared \gls{ddg} node from the scope (since the variable is undeclared before the declaration).
Finally, assignments are either replaced by prefetches, left intact, or dropped completely, based on whether their left- and right-hand sides contain \gls{svm} variables.

\subsubsection{Hardware Requirements}
\label{sec:pf_hw_reqs}

A prefetch load or store is slightly different from a regular memory access: upon a hit in the \gls{tlb}, a prefetch transaction must not be forwarded downstream the memory hierarchy but instead be directly replied by the \gls{iommu} as hit (with don't-care data in case of a load, since data returned by prefetch loads is ignored).
A prefetch that misses in the \gls{tlb} is not different from a regular miss to the hybrid \gls{iommu}: it responds with a miss and drops the transaction.
Conventional \glspl{iommu} cannot support prefetches, since they lack the possibility to drop transactions and the masters using them lack the support for reacting to miss responses.
Whether a load or store is a prefetch can be determined by a single bit sent with the request.
Our implementation uses one bit in the \axi{} \emph{user} field.

\subsection{Multi-Threaded TLB Miss Handling}
\label{sec:impl:par_mht}

In the original implementation of the hybrid \gls{iommu} we are using, \gls{tlb} misses (only metadata) were enqueued by the \gls{iommu} in a hardware queue~\cite{vogel_on_acc_ptw_2017}.
This leftover from conventional \glspl{iommu} presented a centralized bottleneck not required by the design, so we removed it and instead let \glspl{pe} add an entry to a software queue located in the L1 data memory of their cluster upon a \gls{tlb} miss.
This atomic queue supports multiple parallel consumers and producers, and we implemented the atomicity with one enqueue mutex and one dequeue mutex based on the test-and-set functionality of the L1 memory.

For algorithms that make heavy use of \gls{svm} (especially those processing \glspl{lds}), a single \gls{mht} cannot cope with the rate at which \glspl{wt} enqueue misses.
In this case, the rate at which \gls{tlb} misses are handled becomes the bottleneck.
As an \gls{mht} is dominated by memory latency of the page table walking steps, the way to increase the miss handling rate is to let multiple \glspl{mht} work on different pages in parallel.

Two aspects are central for the design of the parallel \glspl{mht}:
(1) given the sequence of misses in the queue, which \gls{mht} handles which miss, and (2) which \gls{mht} modifies which \gls{tlb} entry.

For distributing misses among the \glspl{mht}, the simplest approach would be to let each \gls{mht} dequeue a miss, walk the page table, reconfigure a \gls{tlb} entry, and wake up the \gls{pe} that enqueued the miss.
However, as two subsequent misses frequently go to the same page due to data locality (for an individual \gls{pe}, e.g., with \gls{dma} bursts, but also for multiple \glspl{pe} with shared data), this approach is not effective:
Whenever a \gls{mht} dequeues a miss to a page that another \gls{mht} is already working on, it wastes run time and memory bandwidth on a redundant page table walk, and it wastes \gls{tlb} capacity by setting up a redundant entry.
Ideally, each \gls{mht} would dequeue all misses on the same page, walk the page table, and then wake up all \glspl{pe} waiting for that page.
However, this would require each \gls{mht} to traverse the entire miss queue (which can contain dozens of entries), locking both mutexes while it rearranges the queue without the misses to that page.
This can take hundreds of clock cycles, during which no other \gls{pe} can enqueue or dequeue misses.

Neither wasteful redundant miss handling nor making the miss queue a sequential bottleneck are acceptable, and our design avoids both:
The \glspl{mht} share their state, i.e., which page each \gls{mht} is currently working on and which \glspl{pe} it is going to wake up, through one word per \gls{mht} in the L1 data memory.
When \gls{mht} $ A $ dequeues a miss, it first checks if another \gls{mht} is already working on the same page.
If so, $ A $ tells the other \gls{mht} to also wake up the \gls{pe} that caused the miss $ A $ just dequeued and dequeues another miss.
If no other \gls{mht} works on the same page, $ A $ performs a prefetch memory access to the page to check whether the page has not been mapped since the miss.
If the prefetch misses, $ A $ sets its state to that page and walks the page table.
When $ A $ is done, it reads its state (which may have been updated in the meantime by other \glspl{mht}) and wakes up all assigned \glspl{pe}.

Modifying a \gls{tlb} entry takes two writes because virtual and physical page frame number together are longer than one data word.
To avoid inconsistencies, the \glspl{mht} must thus ensure mutual exclusion when modifying a \gls{tlb} entry.
Any two different \gls{tlb} slots are indepedent, though, so an \gls{mht} should not preclude another from simultaneously modifying a different slot.
As both \gls{tlb} levels are highly associative, \glspl{mht} have multiple options for the placement of each \gls{tlb} entry.
To make effective use of associativity, the \glspl{mht} should agree upon one replacement order per set.
These three requirements can be met by using one atomic counter per \gls{tlb} set, located in memory shared by all \glspl{mht}, which determines the index of the entry to be replaced next in a set.
An \gls{mht} determines the set number from the virtual page frame number, increments the atomic counter of that set, and modifies the entry at the index returned by the counter.
If the number of \glspl{mht} is comparable to the nmuber of entries per set, the \gls{mht} must additionally lock the entry it modifies.

\subsection{Hybrid-IOMMU-Capable DMA Engine}
\label{sec:impl:vmm-capable_dma}

A hybrid \gls{iommu} requires all masters that use it to be capable of tolerating \gls{tlb} misses and keep track of which transactions missed.
The \gls{dma} engine, however, was originally not designed to deal with error responses in a recoverable way and reported a transfer as complete as soon as it had received the final read or write response of the last burst, regardless of whether all bursts were successful or not.
Thus, when a \gls{pe} saw the completion of a transfer it started, it had no way to tell whether all data read or written were valid at the destination.
To guarantee data integrity, all \gls{tlb} entries required for the completion of a transfer (which can touch up to 17 \SI{4}{\kibi\byte} pages) had to be locked before the transfer could be programmed to the \gls{dma} engine and unlocked after it had completed.
As the \gls{tlb} is shared by multiple clusters, this limited the number of \gls{dma} transfers that could be enqueued at a given time and substantially reduced the effective data transfer bandwidth.

If the \gls{dma} engine can keep track of bursts that missed in the \gls{tlb} and restart them after the miss has been handled, \gls{dma} transfers through the hybrid \gls{iommu} can be much more efficient and scalable:
An AXI burst may not cross a page, so each burst requires exactly one \gls{tlb} entry at the instant its request arrives at the \gls{iommu}.
Requests of consecutive bursts can arrive back-to-back at the \gls{iommu}, so multiple \gls{tlb} entries need to be present only for a short time interval for an entire transfer spanning multiple pages to succeed.

To make the \gls{dma} engine compatible with the hybrid \gls{iommu}, we designed and implemented a \emph{retirement buffer} that keeps track of in-flight bursts.
An entry in the buffer contains all metadata required to uniquely identify and reissue a burst: cluster-external and -internal address, length, AXI ID, \acrshort{dma} transfer ID, and whether it was a read or a write.
When the AXI interface of the \gls{dma} engine sends a request, it adds a new entry to the retirement buffer, and when it receives the final response of a burst, it reports the success or failure of the burst with the responded AXI ID to the retirement buffer.

The retirement buffer must keep the order in which bursts were issued (because AXI bursts with the same ID are ordered) and must be able to complete bursts with different AXI IDs in any order.
For these reasons, the retirement buffer can not be a simple FIFO queue.
An alternative would be to have one FIFO queue per AXI ID, but this would waste hardware since every queue would need to have the capacity to store the maximum number of in-flight transfers.

\begin{figure}
  \centering
  \includegraphics[width=\columnwidth]{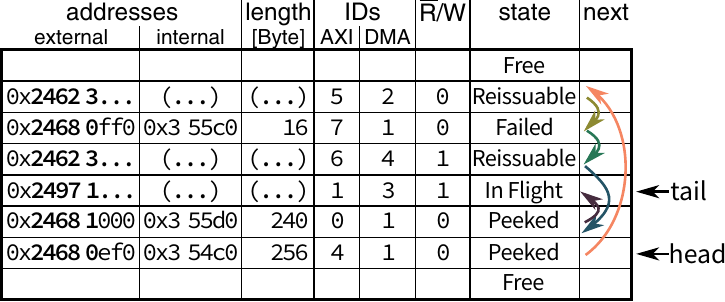}
  \caption{Organization of the burst retirement buffer.}%
  \label{fig:retirement_buffer}
\end{figure}

Instead, our retirement buffer is a linked list implemented in hardware as shown in \cref{fig:retirement_buffer}:
The list entries are stored in a small register file that has as many words as the maximum number of in-flight transfers.
Every word is wide enough to store the burst metadata mentioned above, the state of the entry (free, in-flight, failed, peeked, reissuable), and the index of the next burst entry.
Additionally, the retirement buffer stores the index of the head, where order-preserving peek and pop operations start, and of the tail, where a new in-flight transfer gets enqueued.

The retirement buffer has three main interfaces: one to the AXI transfer unit of the \gls{dma}, one to the internal control unit of the \gls{dma}, and one to the \gls{pe} control interface of the \gls{dma}.
From the transfer unit, the retirement buffer receives commands to add a new in-flight transaction at the tail of the queue, to free a successful transaction, or to mark a transaction as failed.
In the latter two cases, the buffer is traversed from the head, modifying the first non-free transaction that matches the given AXI ID.

To the \gls{dma} control unit, the retirement buffer reports the current number of in-flight and failed bursts and provides the metadata of the next reissuable burst.
When at least one burst has failed, the control unit stops issuing new bursts from its queue and waits for all in-flight bursts to complete.
Once there are no more in-flight bursts, the control unit reissues bursts as soon as they are reissuable until there are no more failed (and in-flight) bursts.
As soon as this is the case, the control unit resumes regular operation by issuing bursts from its queue.

From the \gls{pe} control interface, the cluster-external address of the first (ordered by request, not response) burst with state `failed' can be read and bursts can be marked as reissuable.
For this, a \gls{pe} reads a \gls{dma} register to get the failing external address (or 0 if there is none).
Upon such a read, the retirement buffer marks all `failed' bursts with the same page frame number as `peeked' (so that the same page is not reported twice).
Meanwhile, the \gls{pe} determines the missing physical address and adds it to the \gls{tlb}.
When it is done, it writes the handled virtual address to the same \gls{dma} register, upon which the retirement buffer marks all `failed' or `peeked' bursts with the same page frame number as `reissuable'.
Bursts are then reissued by the control unit in the order of their original requests.

\section{Results}
\label{sec:results}

In this section, we evaluate the performance of our \gls{svm} system implemented on an evaluation platform (\cref{sec:res:eval_platform}) under various conditions (\cref{sec:res:benchmarks}) to demonstrate its significant improvements over the state of the art and identify its limits (\cref{sec:res:synthetic_benchmarks}).
Additionally, we discuss how our hybrid-\gls{iommu}-capable \gls{dma} engine can save a vast amount of hardware buffers compared to conventional \glspl{iommu} and standard \gls{dma} engines (\cref{sec:res:dma_hw_req}).

\subsection{Evaluation Platform}
\label{sec:res:eval_platform}

Our evaluation platform is based on the Xilinx Zynq-7045 \gls{soc}, which features a dual-core ARM Cortex-A9 \gls{cpu}, which we use as host processor, and programmable logic, which we use to implement the cluster-based \gls{pmca} described in~\cref{sec:target_architecture_template}.
The \glspl{pe} within a cluster share \SI{8}{\kibi\byte} L1 instruction cache and \SI{256}{\kibi\byte} tightly-coupled L1 data \gls{spm}, both split into \num{16} banks. %
Ideally, every \gls{pe} can access one 32-bit word in the L1 \gls{spm} per cycle.
Every cluster features a multi-channel \gls{dma} engine that is parametrized to have up to 8 \axi{} read or write bursts in flight at any time, enabling fast movement of data between L1 and L2 memory or shared \gls{dram}.
The \gls{pmca} is attached to the host as a memory-mapped device, interfaced through a kernel-level driver and a user-space runtime.
The host and the \gls{pmca} share \SI{1}{\gibi\byte} of DDR3 \gls{dram}.
The hybrid \gls{iommu} features a two-level \gls{tlb}:
The L1 \gls{tlb} features \num{32} entries, is fully associative, and translates addresses within a single cycle.
The L2 \gls{tlb} features \num{256} entries, is 8-way set associative, and translates addresses within up to 6 cycles.
The \gls{iommu} connects the \gls{pmca} to the \gls{acp} of the Zynq, allowing the \gls{pmca} to access the shared main memory coherent to the data caches of the host.

This platform enables us to study and evaluate the system-level integration of a \gls{pmca} in a \gls{hesoc}.
Thus, we did not optimize the \gls{pmca} for implementation on the \glsknown{fpga}; the \gls{fpga} should be seen as an emulator instead of a fully-optimized accelerator.
We adjusted the clock frequencies of the different components to obtain ratios similar to a real \gls{hesoc} with host and \gls{pmca} running at \SI{2133}{\mega\hertz} and \SI{500}{\mega\hertz}, respectively.
The DDR3 \gls{dram} is clocked at \SI{533}{\mega\hertz}.
Measuring an actual implementation rather than simulating models ensures all aspects and parameters of the evaluated system---including those we did not elaborate in detail in this paper or might have overlooked---are correctly represented in the results.

\subsection{Benchmark Description}
\label{sec:res:benchmarks}

To evaluate the performance of our \gls{svm} system under various conditions including identifying its limits, we have used two entirely different, configurable benchmark applications.
They were obtained by extracting critical phases from real-world applications suitable for implementation on a \gls{hesoc}, and by parametrizing them over a large parameter space.
They exhibit main-memory access patterns representative for various application domains.

\paragraph{Pointer Chasing (PC)}

This benchmark operates on graphs, stored as vertices linked by pointers.
It is representative for wide variety of pointer-chasing applications from the graph processing domain~\cite{guo_graph_processing_2014}.
Prominent examples include breadth-first or shortest path searching, clustering, and Page\-Rank.
Due to the irregular and data-dependent access pattern to shared memory and low locality between references, PC represents a worst-case scenario for a virtual memory subsystem.
However, \gls{svm} is crucial to allow implementations of PC applications at reasonable effort and performance, because offloading a PC application to an accelerator without \gls{svm} requires modifying all pointers in a graph.
In the benchmark, the host builds up a graph and stores its vertices in a single array in main memory.
Every vertex holds the number of successors, a pointer to an array of successor vertex pointers, and a configurable amount of payload data.
At the offload, the host passes a pointer to the vertex array and the number of vertices to the \gls{pmca}.
On the \gls{pmca}, all \glspl{wt} share the work of traversing the vertex array.
For each vertex, a \gls{wt} reads the number of successors and copies the payload data and successor pointers to a buffer in L1 \gls{spm} using \gls{dma} transfers.
The \gls{wt} then performs a configurable number of computation cycles on the payload and writes the payload to all successors in shared main memory again using \gls{dma} transfers.

\paragraph{Stream Processing (SP)}

This benchmark operates on a sequence of data, transferred from and to main memory in regularly strided blocks.
It is representative for applications that work on streams of data, and examples range from simple one-dimensional filtering of audio data, over two- and three-dimensional image and video filters, to tensor operations in neural networks.
In the benchmark, the host allocates one buffer of configurable size for both input and output (to maximize locality) and then passes the pointer to the buffer and the dimensions of the data blocks to the \gls{pmca}.
On the \gls{pmca}, the \glspl{wt} share the work of performing a configurable number of computation cycles on each block.
Both input and output block are double-buffered in L1 \gls{spm}, so that compute and data transfer always overlap.

\subsection{Benchmark Results}
\label{sec:res:synthetic_benchmarks}

In the following plots, we compare the performance of our implementation and the prior \gls{soa} to an ideal \gls{iommu}, which translates every address within a single cycle---an unbiased, although practically unreachable baseline.
The \gls{soa} implementation is from \cite{vogel_on_acc_ptw_2017}, extended to multiple threads on the \gls{pmca}.
For the relative performance on the $y$-axis, higher values are better.
We evaluate different operational intensities on the $x$-axis by changing the number of computation cycles per data as described above.
The operational intensity of an actual program depends both on the algorithm and the hardware executing it, and this sweep over a range of intensities characterizes our \gls{svm} implementation for a given memory access pattern but independent of a very specific program and processing architecture.

\paragraph{Pointer Chasing (PC)}

\begin{figure}
  \centering
  \includegraphics[width=\columnwidth]{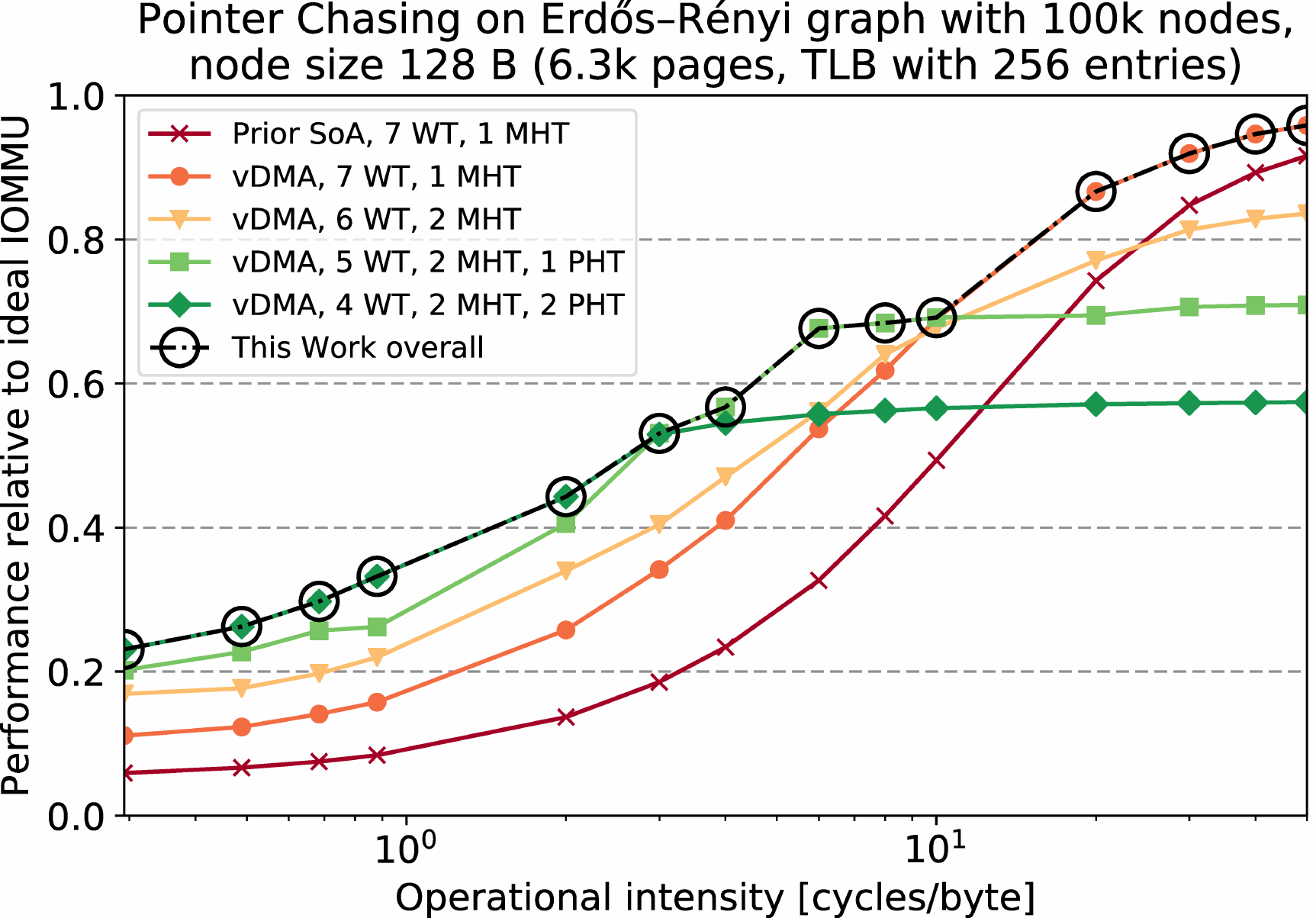}
  \caption{Pointer Chasing (PC) results for different operational intensities.}%
  \label{fig:pointer_chasing}
\end{figure}

\Cref{fig:pointer_chasing} shows the performance of PC normalized to an ideal \gls{iommu} over different operational intensities in cycles per byte.
For example, a single-precision floating-point implementation of the PageRank algorithm has an operational intensity of \SI{1.2}{\cycles\per\byte} given a \gls{fpu} with a divider and around \SI{10}{\cycles\per\byte} for a reduced-precision fixed-point implementations if no \gls{fpu} is available.

In the prior \gls{soa}, the \gls{dma} engine cannot handle \gls{tlb} misses, so the software must ensure the \gls{tlb} entries used by a \gls{dma} transfer are not replaced while that transfer is running.
This locking is the bottleneck of the \gls{soa} implementation (first curve in legend order), and limits its performance to less than \SI{50}{\percent} below \SI{10}{\cycles\per\byte}.
For very high operational intensities, the implementation becomes compute-bound and approaches ideal performance.
Our hybrid-\gls{iommu}-compatible \gls{dma} engine (`vDMA', all other curves) removes that bottleneck.
The second curve is limited by the miss handling throughput of the single \gls{mht} for low operational intensities.
Replacing one of the \gls{wt} with another \gls{mht} (third curve) resolves this bottleneck.
This is effective for low operational intensities, but the missing \gls{wt} reduces performance in the compute-bound limit.
Adding another \gls{mht} (not drawn) does not further improve performance because two \glspl{mht} are sufficient to handle the misses caused by six \glspl{wt}.
Instead, we replace one of the \glspl{wt} by a \gls{pht} (fourth curve), which causes \gls{tlb} entries to be set up ahead of the instant the \glspl{wt} need them.
This is very effective in the memory-bound case, increasing performance by another \SIrange{20}{30}{\percent}.
The fourth curve is now prefetch-limited: the single \gls{pht} cannot always prefetch early enough, because the \gls{pht} itself needs to dereference pointers to determine prefetch targets.
Any dereference that causes a \gls{tlb} miss will block the \gls{pht} until the miss is resolved.
Thus, replacing another \gls{wt} with an \gls{mht} helps increasing performance in memory-bound cases by an additional \SI{20}{\percent}.

Depending on the operational intensity, one of the configurations is optimal.
However, as \glspl{mht} and \glspl{pht} can be inserted in software, e.g., at compile time based on profiling runs or even at run time for largely varying operational intensities, our work significantly improves performance for \emph{all} operational intensities by making optimal use of \glspl{pe}.
The last curve shows the overall optimum configuration.
For crucial operational intensities around \SI{1}{\cycles\per\byte}, our work improves performance by 4\,x compared to the \gls{soa}.
For common intensities (arguably below \SI{10}{\cycles\per\byte}), our work improves the \gls{soa} performance by at least \SI{50}{\percent}.

\paragraph{Stream Processing (SP)}

\begin{figure}
  \centering
  \includegraphics[width=\columnwidth]{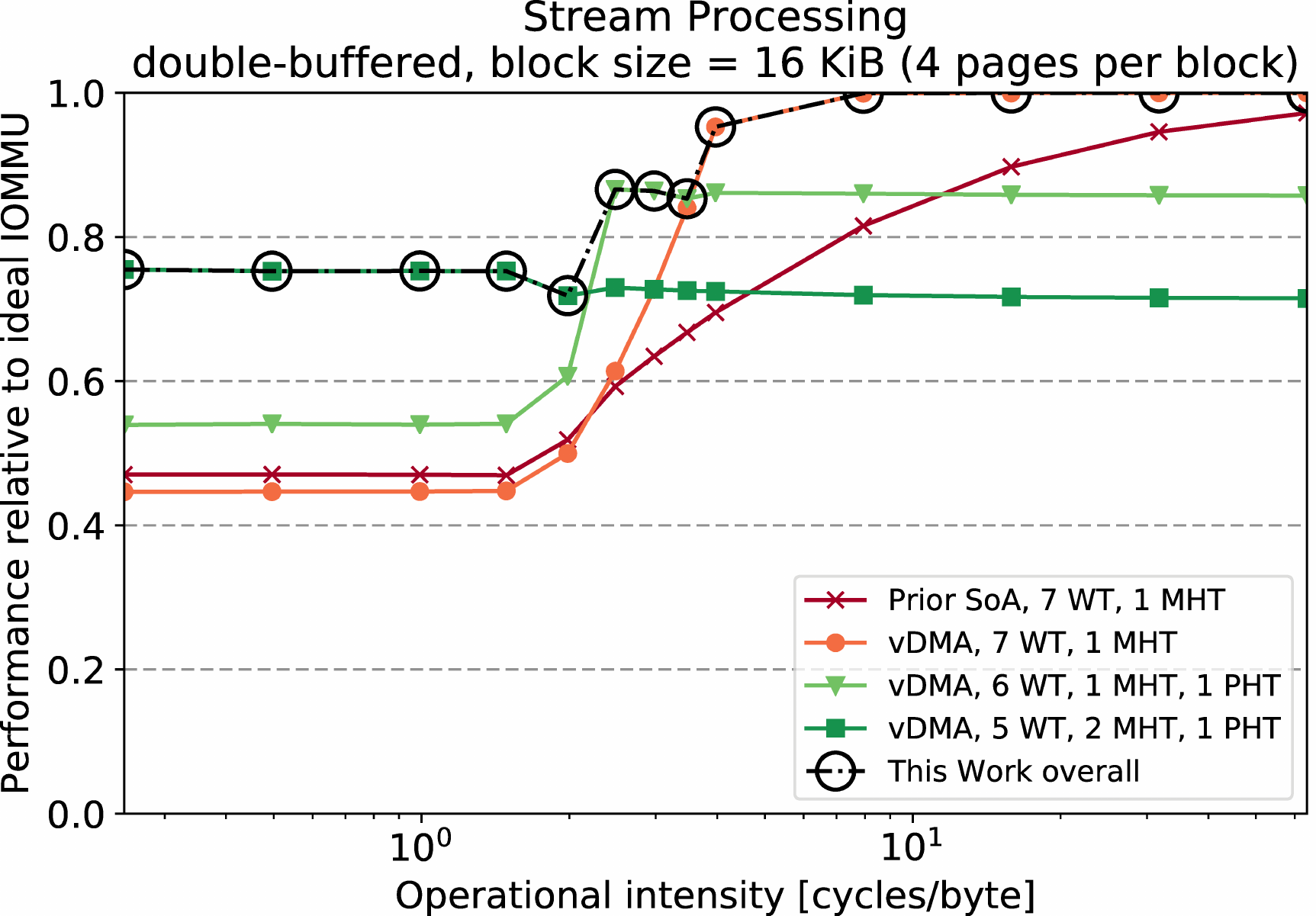}
  \caption{Stream Processing (SP) results for different operational intensities.}%
  \label{fig:stream_processing}
\end{figure}

\Cref{fig:stream_processing} shows the performance of SP normalized to an ideal \gls{iommu} over different operational intensities.
For example, a one-dimensional FIR filter with $N$ coefficients requires $N/2$ \glspl{mac} per transferred data word (each word transferred once in, once out), and a matrix-matrix multiplication requires $1$ \gls{mac} per transferred data word for large matrices.
Assuming \glspl{mac} on the data format are natively supported by the \gls{pmca}, the \gls{pmca} may compute tens, hundreds, or even thousands of \glspl{mac} per cycle, depending on the number and architecture of its parallel \glspl{pe}.
Thus, stream processing kernel-architecture combinations may be found anywhere on the $x$-axis of \Cref{fig:stream_processing}.

In the prior \gls{soa} (first curve in legend order), a \gls{wt} setting up a \gls{dma} transfer ensures that no \gls{tlb} misses occur during the transfer by explicitly setting up \gls{tlb} entries and locking them during the transfer.
For memory-bound kernels, this is slightly more performant than handling misses by the hybrid-\gls{iommu}-compatible \gls{dma} engine (second curve), because the latter stalls on every miss.
(If that performance difference was larger, \gls{tlb} entries could be set up in advance also for the vDMA.
In contrast to the prior \gls{soa}, where locks on \gls{tlb} entries had to be coded manually to avoid deadlocks, instructions for setting up \gls{tlb} entries in advance for the vDMA can be inserted automatically at compile time.)
For a range of more compute-intensive kernels, however, this locking is the bottleneck of the \gls{soa} implementation, and removing it improves performance by up to \SI{35}{\percent}.
When only few cycles are executed per transferred byte, performance is dominated by the memory latency in handling \gls{tlb} misses.
Adding another \gls{mht} (not drawn) does not change this, because only the input data stream requires one new page at a time.
Instead, we replace one of the \glspl{wt} with a \gls{pht} (third curve), which increases performance by \SIrange{20}{40}{\percent} in the memory-bound case.
This configuration is limited by the \emph{throughput} of the \gls{mht}, and because the \gls{pht} causes more than one page to be outstanding in the miss queue, there is now work for another \gls{mht}.
Indeed, adding another \gls{mht} (fourth curve) increases performance by another \SI{40}{\percent}, up to the point where it limited by the throughput of the \gls{pht} in the memory-bound case and by the five \glspl{wt} in the compute-bound case.
Adding another \gls{pht} might increase performance even further, but the current \gls{pht} generation algorithm does not support distributing the prefetches for a single memory access stream among two \glspl{pht}.

The last curve shows the overall optimum of all configurations of our work.
Our work improves performance compared to the \gls{soa} by up to \SI{60}{\percent} for memory-intense kernels, and reduces the overhead compared to an ideal \gls{iommu} to below \SI{25}{\percent} for \emph{any} operational intensity.
Our work also reduces the operational intensity at which that overhead is below \SI{10}{\percent} to ca.\ \SI{4}{\cycles\per\byte}.
As the optimal configuration again can be selected at compile time or even changed at run time, our work significantly improves performance over the full spectrum of operational intensities also for very regular memory access patterns by replacing \glspl{wt} with \glspl{mht} or \glspl{pht} when it improves overall performance.

\subsection{Hardware Requirements of Hybrid-IOMMU-Capable DMA}
\label{sec:res:dma_hw_req}

Making the \gls{dma} engine compatible with the hybrid \gls{iommu} not only improves performance compared to the \gls{soa}, it also dramatically reduces the amount of memory required to buffer \gls{dma} bursts that miss in the \gls{tlb}.
Our DMA engine is parametrized to have up to 8 \axi{} read or write bursts in flight at any time.
Each burst can transfer up to \SI{2}{\kibi\byte}. %
The total maximum amount of data in flight is \SI{16}{\kibi\byte}, and a buffer of this size would be required to enable other masters to continue accessing \gls{svm} in the worst case scenario where 8 write bursts miss in the \gls{tlb}.
The retirement buffer in our \gls{dma} engine stores just the metadata of each burst:
32 bit for the virtual start address, \SI{16}{\bit} for the local start address, \SI{3}{\bit} for the ID, \SI{8}{\bit} for the length of the burst, and \SI{3}{\bit} for the status of the burst; less than \SI{8}{\byte} in total.
Thus, the retirement buffer requires just \SI{64}{\byte} for the same \gls{tlb} miss tolerance as the \SI{16}{\kibi\byte} data buffer---a factor 256 less.

\section{Conclusion}
\label{sec:conclusion}

In this work, we presented and evaluated our scalable and efficient \gls{svm} solution for \glspl{hesoc}.
It is based on a hybrid \gls{iommu} and advances the state of the art in three important ways:
First, compiler-generated \glspl{pht} proactively fill the \gls{tlb} to minimize the rate of \gls{tlb} misses.
Second, a variable number of parallel \glspl{pht} handle \gls{tlb} misses to scale the miss handling throughput with the demand.
Third, a hybrid-\gls{iommu}-capable \gls{dma} engine supports parallel burst \gls{dma} transfers to \gls{svm} without additional buffers.
Compared to the state of the art, our work improves \gls{pmca} performance for memory-intensive kernels by up to 4$\times$ for irregular and by up to \SI{60}{\percent} for regular memory access patterns.
Compared to using data buffers to absorb bursts from \gls{dma} engines in a conventional \gls{iommu}, our solution requires two orders of magnitude less memory and scales better, as it only stalls the missing \gls{dma} engine.
In the future, we plan to explore compiler-generated \glspl{pht} for kernels that mandate speculative prefetching, improve per-thread miss handling throughput by supporting out-of-order page table walking, and avoid stalling the entire \gls{dma} on a \gls{tlb} miss while maintaining memory order guarantees.

\section*{Acknowledgments}

\thanks{The authors thank J.~Weinbuch for his work on multi-threaded \acrshort{tlb} miss handling during his Master's Thesis.}

\bibliographystyle{IEEEtran}
\bibliography{paper}

\end{document}